\documentclass[aps,prd,amssymb,amsmath,amsfonts,superscriptaddress,nofootinbib,eqsecnum,reprint,showpacs,longbibliography]{revtex4-2}
\usepackage{graphicx}
\usepackage{bm}
\usepackage{color}
\usepackage{commath}
\usepackage{dcolumn}% Align table columns on decimal point
\usepackage{xcolor}
\usepackage[unicode,colorlinks=true,citecolor=linkcolor,linkcolor=linkcolor,urlcolor=linkcolor]{hyperref}
\usepackage{cleveref}
\usepackage{orcidlink}
\usepackage{hyperref}
\hypersetup{colorlinks=true,linkcolor=magenta,anchorcolor=green,citecolor=black,filecolor=black,menucolor=black,urlcolor=black}
\usepackage{tensor}
\newcommand{\bb}{\bar{b}}
\newcommand{\la}{\langle}
\newcommand{\ra}{\rangle}
\begin{document}
\title{A Non-linear Massive Gravity Theory of Geometric Origin}
\author{Thibault Damour}
 \email{damour@ihes.fr}
 \affiliation{%
 Institut des Hautes Etudes Scientifiques, 91440 Bures-sur-Yvette, France
}%

\author{Tamanna Jain}%
 \email{tj317@cam.ac.uk}
\affiliation{%
LPENS, Département de physique, Ecole normale supérieure, Université PSL, Sorbonne Université, Université Paris Cité, CNRS, 75005 Paris
}%
  \affiliation{%
 Institut des Hautes Etudes Scientifiques, 91440 Bures-sur-Yvette, France
}%
    \affiliation{%
 Department of Applied Mathematics and Theoretical Physics,University of Cambridge, Wilberforce Road CB3 0WA Cambridge, United Kingdom
}%

\date{\today}

\begin{abstract}
    We study the number of propagating degrees of freedom, at non-linear order, in  torsion gravity theories, a class of modified theories of gravity that include a propagating torsion in addition to the metric.  We focus on a three-parameter subfamily of theories (``torsion bigravity") that contains,  at linear order, only two physical excitations: a massless spin-2 one (with two degrees of freedom) and a massive spin-2 one (with five degrees of freedom).  We study the dynamics of the massive spin-2 field in the limit where the torsion field decouples from the metric. The number of degrees of freedom of the torsion field is found to {\it change, at non-linear order, from five to nine}. 
\end{abstract}

\maketitle

\section{Introduction}
The predictions of the theory of General Relativity (GR) have been  so far  in outstanding agreement with gravitational observations and experiments over a very large range of scales, and field strengths (for reviews, see, e.g., chapter 21 in \cite{ParticleDataGroup:2024cfk} or the book \cite {Will:2018bme}). Nevertheless, since its inception, there has been a constant effort to study possible extensions of GR. The search for alternatives to GR is motivated by open questions both at ultraviolet scales (e.g. \cite{Kiefer:2023bld})
 and at infrared scales (e.g. \cite{Joyce:2014kja, Frusciante:2019xia, Salucci:2020eqo}). The existence of consistent alternative theories of gravity is also useful for suggesting tests of GR \cite{Berti:2015itd}.

In the geometric approach to gravity, modifying  GR can stem from extending the geometrical structure of spacetime. As early as 1923, Cartan \cite{Cartan:1923zea,Cartan:1924yea} suggested to extend GR by formally keeping the Einstein-Hilbert action, but in a first-order formalism, considering the metric, and a general (metric-preserving) {\it torsionfull} connection, as independent fields (``Einstein-Cartan theory"). Cartan imagined that the torsion could be sourced by some material spin density.  This idea was later made concrete by Weyl \cite{Weyl:1950xa} who studied the coupling of (Dirac) spinor fields to gravity within a first-order formalism and pointed out that, in this approach, the  spinors provide a spin-density source $\tau^{\mu\nu\lambda}$ for the torsion proportional to 
$\bar \psi \gamma^{[\mu} \gamma^\nu \gamma^{\lambda]} \psi$. He also pointed out that in Einstein-Cartan theory the torsion does not propagate (i.e. is not a dynamical field), but is constrained to be proportional to its source $\tau^{\mu\nu\lambda}$, thereby leading (in a second-order formalism for the metric) to a contact term proportional to $\tau^{\mu\nu\lambda} \tau_{\mu\nu\lambda}$.
 The ideas of Cartan and Weyl were further extended and developed by Sciama, %\cite{Sciama:1962bnm,Sciama:1964wt},
 Kibble, % \cite{Kibble:1961ba} 
 and  others (see, e.g., \cite{Hehl:1976kj} for a review).

In this work, we study a class of geometric theories of gravitation, introduced in Refs.~\cite{Sezgin:1979zf,Hayashi:1979wj,Hayashi:1980qp,Sezgin:1981xs},
    which generalize the Einstein-Cartan theory in having a {\it dynamical} torsion field that propagates away from its source
$\tau^{\mu\nu\lambda}$. We focus on the theories defined in Refs.~\cite{Sezgin:1979zf,Hayashi:1979wj,Hayashi:1980qp,Sezgin:1981xs} that contain (around flat-space backgrounds), in addition to  Einsteinlike massless spin-$2$ excitations,  (even-parity) massive spin-$2$ excitations (embodied in the torsion field) as well as (pseudo-scalar) massive spin-$0$ excitations . Such theories will hereafter be referred to as {\it torsion gravity} theories. Below, we shall focus on the subclass of torsion gravity theories, called
{\it torsion bigravity} theories, which 
contain only massive and massless spin-$2$ excitations.

Torsion gravity theories were recently revived with the hope that the geometrically-rooted massive spin-2 field they contain might define a   theoretically healthy, and phenomenologically interesting,  modification of gravity both in cosmological situations \cite{Nair:2008yh,Nikiforova:2009qr,Deffayet:2011uk,Nikiforova:2018pdk}, and in strong-field regimes \cite{Damour:2019oru,Nikiforova:2020fbz,Nikiforova:2021xcj,Nikiforova:2022pqy}. The hope was that  a geometrically well-motivated  origin  for massive spin-$2$ excitations (through a dynamical torsion) would alleviate the problematic issues that arise when the extra massive spin-$2$ field is linked to the addition of a second metric tensor, $f_{\mu\nu}$, besides $g_{\mu\nu}$, as in generic {\it bimetric gravity} theories. Let us recall that there are several problematic issues associated with the usual (Fierz-Pauli-type \cite{Fierz:1939ix}) descriptions of massive spin-2 fields, notably: the Van-Dam-Veltman discontinuity \cite{vanDam:1970vg}, the Vainshtein radius issue \cite{Vainshtein:1972sx}, and the additional, ghostlike, sixth degree of freedom (dof)  \cite{Boulware:1972yco}.

Generic bimetric gravity theories (see, e.g., Ref. \cite{Damour:2002ws}), involving some general potential ${\mathcal V}(g_{\mu\nu}, f_{\mu\nu})$, suffer from the arising, at the nonlinear level, of the unwanted, additional, Boulware-Deser ghostlike dof,
see, e.g., \cite{Babichev:2009us}. The Boulware-Deser ghost was avoided by generalizing the special potential of the ``ghost-free" massive gravity theory of Ref.~\cite{deRham:2010kj} to a bimetric gravity framework \cite{Hassan:2011zd}.
Note, however, that recent works (see \cite{Bellazzini:2023nqj} and references therein) suggest that the theory of massive spin-$2$ excitations described by the ghost-free action of Ref.~\cite{deRham:2010kj} is only valid up to an energy scale which cannot be parametrically larger than the mass of the spin-2 field.

By contrast with bimetric gravity theories, 
it was shown in Refs.~\cite{Nair:2008yh,Nikiforova:2009qr,Deffayet:2011uk,Nikiforova:2018pdk,Damour:2019oru,Nikiforova:2020fbz,Nikiforova:2021xcj,Nikiforova:2022pqy} that torsion gravity theories have the good feature of  keeping the same number of dof in many situations,
involving, notably, dynamical (but linear) torsion excitations around generic nonlinear Einstein backgounds, or fully nonlinear spherically symmetric (but stationary) torsion gravity solutions. In addition, it was also found \cite{Nikiforova:2020fbz} that the massive spin-$2$ torsion excitation does not introduce a
Vainshtein radius in static, spherically-symmetric
nonlinear solutions.
On the other hand, the study of time-dependent torsion perturbations
  suggested that torsion gravity theories are only valid up to curvature scales of order of the squared spin-2 mass \cite{Nikiforova:2021xcj,Nikiforova:2022pqy}.]

Motivated by the good features of torsion gravity
theories found in previous studies,
the aim of the present work is to investigate the number of propagating dof of torsion gravity theories in situations which have not yet been considered. Namely, we consider (vacuum) solutions which are both time-dependent and nonlinear in the torsion field. To simplify our investigation, we focus on the non-linear dynamics of the massive spin-2 (torsion) excitation in a well-defined  limit where it decouples from the massless spin-2 (metric) one. 

The paper is organized as follows. In Sec.~\ref{Sec:FormalismTG}, we review the formalism and present the field equations for general torsion gravity theories along with reviewing the specialized three-parameter subfamily of torsion bigravity theories. The decoupling limit of the massive spin-2 field from the massless spin-2 field is presented in Sec.~\ref{Sec:DecoupleLimit}. We first demonstrate the general, all-order structure of the field equations in the decoupling limit and then explicitly derive the field equations and investigate the constraints (degrees of freedom) at the linear and the next-to-linear order in Sec.~\ref{Sec:FieldEquationsTG}. To explicitly understand the number of degrees of freedom for the field equations, we present several alternative action formulations of the decoupling limit in Sec.~\ref{Sec:ActionformTG}. We then investigate the general structure of the constraints  in a resulting first-order (Hamiltonian)  formalism in Sec.~\ref{Sec:Structureaction}. The number of dof is derived at the linear level in Sec.~\ref{Sec:TGLinear} and, finally, at the non-linear level in Sec.~\ref{Sec:TGnon-linear}. Brief conclusions are presented in Sec.~\ref{Conclusions}.

\section{Formalism of Torsion Gravity}
\label{Sec:FormalismTG}
In this section, we discuss the formalism of torsion gravity theories following the notations of Ref.~\cite{Nikiforova:2018pdk}, as updated in Ref.~\cite{Damour:2019oru}. 
The Latin indices $i,j...=0,1,2,3$ are used to denote vierbein indices linked to a local orthonormal frame ($e_i^{\mu}$) while Greek indices $\mu,\nu...=0,1,2,3$ are used to denote the space-time indices related to the coordinate frame $x^{\mu}$. 
The frame indices are moved by the Minkowski metric $\eta_{ij}$ whereas the space-time indices are moved by the metric $g_{\mu\nu}=\eta_{ij}e^i_{\mu}e^j_{\nu}$, where $e^i_{\mu}$ is the co-frame, such that $e^i_{\mu} e_j^{\mu}= \delta^i_j $ . The metric signature used is $(-,+,+,+)$.

The most general action with propagating torsion (expressed in terms of the co-frame $e^i_{\mu}$ and a general SO(3,1) connection $A_{ij \mu}$)  having as spectrum a massless spin-2 field, a massive spin-2 field, and a massive pseudo-scalar field,  is
$S_{\mathrm{TG}} +  S_{\rm matter}$, where the field part of the action reads
\begin{align}
\label{eqn:STGB}
    S_{\mathrm{TG}}[e^i_{\mu},A_{ijk}]=\int d^4x \sqrt{-g} \mathcal{L}_{TG}[e,\partial e,\partial^2 e,A,\partial A]~,
\end{align}
with
\begin{align}
\label{eqn:LTG}
    \mathcal{L}_{TG} &= c_{R} R + c_{F} F +c_2+ c_3F^{ij}F_{ij}+c_4F^{ij}F_{ji} \nonumber \\
   & -\frac{1}{3}(c_3+c_4)F^2+c_6(\epsilon\cdot F)^2 \nonumber \\
    &= c_{R} R + c_{F} F +c_2 +  c_{F^2} \left(F_{(ij)}F^{(ij)}-\frac{1}{3}F^2\right)\nonumber \\
    &+c_{34}F_{[ij]}F^{[ij]}+c_6 \left(\epsilon\cdot F\right)^2~.
    \end{align}
The second form of the latter action uses 
the symmetric ($F_{(ij)}$) and antisymmetric  ($F_{[ij]}$) parts of $F_{ij}= F_{(ij)}+ F_{[ij]}$, and the notation
    \begin{align}
    \label{eqn:LTG1}
    c_{34}& \equiv c_3-c_4~,\hspace{0.5cm}
    c_{F^2} \equiv c_3+c_4~.
    \end{align}

Here, $c_{R}$,  $c_{F}$, $c_2$, $c_{3}$,  $c_{4}$, and $c_6$ are coupling constants, $R$ is the usual Einsteinian Ricci scalar of the metric $g_{\mu \nu}$, while $F_{ijkl}$, $F_{ij}$,  and $F$ are the (frame components of the) curvature tensor, Ricci tensor ($F_{ij}= \eta^{kl} F_{k i l j}$) and curvature scalar ($ F=\eta^{ij} F_{ij}$) defined by the (torsionfull) connection $A_{ij\mu}$. The metric-preserving character of
the SO(3,1) connection $A_{ijk} \equiv A_{ij\mu} e^\mu_k$ is encapsulated in the algebraic fact that  $A_{ijk}= - A_{jik}$.
The term  $\epsilon\cdot F \equiv
\epsilon_{ijkl}F^{ijkl}$
involves the Levi-Civita antisymmetric tensor $\epsilon_{ijkl}$. 
The explicit expression of the curvature tensor $F_{ijkl}$ and its corresponding avatars $F_{ij}$ and $F$ are,
\begin{align}
    F_{ijkl}=&e^{\mu}_ke^{\nu}_l\left(\partial_{\mu}A_{ij\nu}-\partial_{\nu}A_{ij\mu}\right.\nonumber\\
    &\left.+\eta^{mn}A_{im\mu}A_{nj\nu}-\eta^{mn}A_{im\nu}A_{nj\mu}\right)~,\\
    F_{ij}=& \eta^{kl}F_{ikjl}~,\hspace{0.8cm} F=\eta^{ij}F_{ij}~.
\end{align}
The curvature tensor $F_{ijkl}$, like the metric Riemann tensor $R_{ijkl}$, is anti-symmetric under the exchanges $i\leftrightarrow j$ and $k\leftrightarrow l$, however, contrary to $R_{ijkl}$, it is not symmetric under the exchange $ij\leftrightarrow kl$.

The torsion-gravity field equation obtained by  varying the vierbein $e^i_{\mu}$ in $S_{\mathrm{TG}} +  S_{\rm matter}$ reads
\begin{align}
\label{eqn:EOMemu}
     & c_{R}\left(R_{ij}-\frac{1}{2}\eta_{ij}R\right)+c_{F}\left(F_{ij}-\frac{1}{2}\eta_{ij}F\right)-\frac{1}{2}c_2\eta_{ij}\nonumber\\
    &+c_3(F^{k}_{i}F_{kj}+F^{kl}F_{kilj})+c_4(F_{i}^{k}F_{kj}+F^{lk}F_{kilj})\nonumber \\
    &-\frac{2}{3}(c_3+c_4)F F_{ij}+2c_6 \epsilon^{klmi}F_{klmj}(\epsilon\cdot F)-\frac{1}{2}\eta_{ij}L^{(2)}=\mathcal{T}_{ij}~,
\end{align}
where $\mathcal{T}_{ij}$ is the gravitational matter source, and
\begin{align}
    L^{(2)}=c_3F^{ij}F_{ij}+c_4F^{ij}F_{ji}-\frac{1}{3}(c_3+c_4)F^2+c_6(\epsilon\cdot F)^2~.
\end{align}
The field equation corresponding to varying the connection $A_{ijk}$ reads
\begin{align}
\label{eqn:EOMA}
    &\left[\eta_{ik}\left(D^m P_{jm}-\frac{2}{3}D_jP\right)-D_i P_{jk}\right]\nonumber \\
    &-\left[\eta_{jk}\left(D^mP_{im}-\frac{2}{3}D_i P\right)-D_jP_{ik}\right]\nonumber \\
    &+4c_6\epsilon_{ijkm}D^m(\epsilon\cdot F)+H_{ijk}+S_{ijk}= \mathcal{T}_{ijk}~,
\end{align}
where $\mathcal{T}_{ijk}$ is the material torsion source, and where
\begin{align}
\label{eqn:Hijk}
    H_{ijk}&= c_{F}\left(K_{ikj}-K_{jki}-K_{il}^{l}\eta_{jk}+K_{jl}^{l}\eta_{ik}\right)~,\\
    \label{eqn:Sijk}
    S_{ijk}&=\frac{1}{c_{F}}H_{mnk}\left(\eta_{im}P_{jn}-\eta_{jm}P_{in}-\frac{2}{3}\eta_{im}\eta_{jn}P\right.\nonumber\\
    &\left.+2c_6\epsilon_{ijmn}(\epsilon\cdot F)\right)~,\\
    \label{eqn:Pij}
    P_{ij}&=c_3 F_{ij}+c_4F_{ji},\hspace{0.5cm}P\equiv\eta_{ij}P_{ij}~.
\end{align}
Here, $D_i$ denotes the covariant derivative with respect to the connection $A_{ijk}$: e.g., $D_i V^j=e_i^{\mu}(\partial_{\mu}V^j+A^j_{\; \, k\mu}V^k)$;
 and $K_{ijk}$ is the contorsion tensor, i.e the difference between  the connection $A_{ijk}$ and the metric-induced (Levi-Civita) connection $\omega_{ijk} \equiv \omega_{ij \mu} e^\mu_k$:
\begin{align}
    K_{ijk}=A_{ijk}-\omega_{ijk}~.
\end{align}

The field equations \eqref{eqn:EOMemu} and \eqref{eqn:EOMA} of torsion gravity generally depend on the six coupling parameters: $c_{F}$, $c_{R}$, $c_2$, $c_{3}$, $c_{4}$, and $c_6$. 

In the present work, we follow Ref.~\cite{Damour:2019oru} in focussing on the 3-parameter subfamily of torsion bigravity theories where
\begin{align}
\label{eqn:assump1}
    c_2 &= 0~,\hspace{0.6cm} c_3 = c_4~,\hspace{0.3cm}\mathrm{and}\hspace{0.3cm} c_6 =0~,
\end{align}
such that, in the notation of Eq.~\eqref{eqn:LTG1}, we have
\begin{align}
    c_{34}&= 0~,\hspace{0.6cm} c_3 = c_4 = \frac{1}{2}c_{F^2}.
\end{align}
These theories  have only a massless spin-2 and a massive spin-2 in their spectrum.
The corresponding torsion bigravity Lagrangian takes the simple form 
\begin{align}
\label{eqn:Lagnew}
    \mathcal{L}_{TBG} &= c_{R} R + c_{F} F + c_{F^2} \left(F_{(ij)}F^{(ij)}-\frac{1}{3}F^2\right)~.
\end{align}

The corresponding field equations simplify. In particular, Eq.~\eqref{eqn:EOMA} now reads
\begin{align}
\label{eqn:EOMAnew}
    & \left[\eta_{ik}\left(D_m P_{jm}-\frac{2}{3}D_jP\right)-D_i P_{jk}\right]\nonumber \\
    &-\left[\eta_{jk}\left(D_mP_{im}-\frac{2}{3}D_i P\right)-D_jP_{ik}\right]\nonumber \\
    &+H_{ijk}+S_{ijk}= \mathcal{T}_{ijk},
\end{align}
where $P_{ij} = c_{F^2} F_{(ij)}$, and $H_{ijk}$ and $S_{ijk}$ are as defined above (with $c_6=0$). Note that the latter field equation only involves the symmetric part $F_{(ij)}$ of $F_{ij}$. 
In this work, we focus on vacuum field equations, in absence of material sources (i.e. with $\mathcal{T}_{ij}=0$ and $\mathcal{T}_{ijk}=0$).

Our restriction to a subfamily of torsion gravity theories will be of sufficient generality for our purpose because we shall show below that the subfamily \eqref{eqn:Lagnew}, which showed healthy properties in previously considered situations, becomes unhealthy when considered at the nonlinear level. 

\section{Decoupling gravity from the torsion dynamics}
\label{Sec:DecoupleLimit}
It will be convenient here to further simplify our task by focusing on the nonlinear dynamics of  the massive spin-2 excitation contained in the connection $A_{ijk}$ by considering a limit where the dynamics of $A_{ijk}$ decouples from the dynamics of the metric $g_{\mu \nu}$. [Here, we assume that the vierbein $e^i_{\mu}$ has been
(locally) algebraically expressed in terms of the metric.]
To this effect let us write the metric as 
\begin{equation} \label{gexp}
g_{\mu \nu}= \eta_{\mu \nu} + \frac{1}{\sqrt{c_R}} {\hat h}_{\mu \nu}.
\end{equation}
Inserting the latter parametrization of the metric in the Einsteinian part of the action density $ L_g= \sqrt{-g} c_{R} R(g)$ yields (after an integration by parts) an action density of the form
\begin{equation}
L_g= (\partial {\hat h})^2 + \frac{1}{\sqrt{c_R}} {\hat h}(\partial {\hat h})^2 + \frac{1}{c_R}{\hat h}^2 (\partial {\hat h})^2+ \cdots
\end{equation}
Therefore, in the limit 
$c_{R} \rightarrow {\infty}$, the metric perturbation ${\hat h}_{\mu \nu} $ (which also couples to the matter and to other fields proportionally to
$1/\sqrt{c_R}$) decouples as a free massless spin-2 excitation. 

As a consequence, in the limit $c_{R} \rightarrow {\infty}$ (henceforth referred to as the {\it decoupling limit}), the dynamics of the connection $A_{ijk}$  (equal in this limit to the
contorsion tensor $K_{ijk}$ because $g_{\mu \nu}= \eta_{\mu \nu} + \mathcal{O}\left(\frac{1}{\sqrt{c_R}} \right)$, $e^i_{\mu}=\delta^i_{\mu} 
 + \mathcal{O}\left(\frac{1}{\sqrt{c_R}} \right)$, and $\omega_{ijk} = \mathcal{O}\left(\frac{1}{\sqrt{c_R}}\right)$) takes place within the flat background spacetime
$\eta_{\mu \nu}$ (endowed with a flat vierbein field $e^i_{\mu}=\delta^i_{\mu}$ with corresponding Lorentzian coordinates $x^i=x^\mu$). 

Below, we study the resulting decoupled (and  sourceless) field equations for $A_{ijk}$. The crucial point for our study is that we are retaining in this limit all the nonlinearities in $A_{ijk}$ entailed by the  action
\begin{align}
\label{eqn:Lagdec}
    \mathcal{L}^{\rm decoupled}_{TBG}(A) =& c_{F} F(A) \nonumber\\
    &+ c_{F^2} \left(F_{(ij)}(A)F^{(ij)}(A)-\frac{1}{3}(F(A))^2\right)~,
\end{align}
where $F(A) \sim \partial A + A A$, namely
\begin{align}
    F_{jl} =& \eta^{ik}\left(\partial_{k}A_{ijl}-\partial_{l}A_{ijk}+\eta^{mn}A_{imk}A_{njl}\right.\nonumber\\
    &\left.-\eta^{mn}A_{iml}A_{njk}\right)~,\\ \nonumber
    F=& \eta^{jl}F_{jl}~.
\end{align}
Here, all partial derivatives are with respect to
the Lorentzian coordinates $x^i$.

The action $\mathcal{L}^{\rm decoupled}_{TBG}(A)$ is {\it quartic} in $A_{ijk}$. It has a formal similarity with a Yang-Mills action, though it differs from what would be the normal 
Yang-Mills action for a $SO(3,1)$ connection.

\section{Structure of the equations of motion of torsion bigravity in the decoupled limit}
\label{Sec:FieldEquationsTG}
A direct derivation of the field equations from the action \eqref{eqn:Lagdec} would yield second-order equations for the twenty-four components of the connection $A_{ijk}= - A_{jik}$. However, following Refs.~\cite{Nikiforova:2009qr,Deffayet:2011uk}, the field equations for $A_{ijk}$ can be conveniently re-expressed as field equations for the symmetric part  $F_{(ij)}$  of the Ricci tensor of $A_{ijk}$. 

Indeed, taking into account that $D \sim \partial + A$, 
$P_{ij} \sim c_{F^2} F_{(ij)}$, $H_{ijk} \sim c_F A_{ijk}$,
and $S_{ijk} \sim \frac{1}{c_F} A_{ijk} \cdot F_{(lm)}$, the torsion field equations Eq.~\eqref{eqn:EOMAnew} have the following structure (indicating only the number of indices)
\begin{equation} 
\label{eom1}
  c_{F^2}  ( \partial_1 + A_3) F_{(2)}+ c_F A_3 + c_{F^2} A_3 F_{(2)} = 0~.
\end{equation}
Let us introduce the notation
\begin{equation} \label{kappa}
    \kappa^2 \equiv \frac{c_F}{c_{F^2}},
\end{equation}
to denote the squared mass of the massive spin-2 excitation. [More precisely, Eq.~\eqref{kappa} defines the squared mass in the limit where $\frac{c_F}{c_R} \to 0$. When $c_R$ is finite, we would have $ \kappa^2 = \frac{c_F}{c_{F^2}}\left(1+ \frac{c_F}{c_R}  \right)$.] The structure of the above field equation then reads
\begin{equation}
\label{AvsF0}
  \partial_1   F_{(2)} + (\kappa^2 + F_{(2)}) A_3=0\,.
\end{equation}
This equation can be formally solved in terms of $A_3$ to yield
\begin{equation} \label{AvsF}
 A_3= -(\kappa^2 + F_{(2)})^{-1} \partial_1   F_{(2)} ~,
\end{equation}
where the formal inverse $(\kappa^2 + F_{(2)})^{-1}$ is perturbatively defined as an all-order expansion in powers of $\frac{F_{(2)}}{\kappa^2}$. We shall not study here the singularities that might appear in this inverse when considering strong fields $\frac{F_{(2)}}{\kappa^2} = \mathcal{O}(1)$ [The occurrence of similar singularities, leading to a lower bound on the mass $\kappa$, were pointed out in Refs.~\cite{Nikiforova:2021xcj,Nikiforova:2022pqy}.].

Formally inserting the expression Eq.~\eqref{AvsF} for $A_3$ in the definition
$F_{(2)} \sim \partial A_3 + A_3 A_3 $ of $F_{(2)}$ then leads to a nonlinear (Fierz-Pauli-like \cite{Deffayet:2011uk}) second-order equation for the dimensionless
symmetric tensor field 
\begin{equation}
    {\hat F}_{(ij)} \equiv \frac{{ F}_{(ij)}}{\kappa^2}~,
\end{equation}
 having the symbolic structure
 \begin{equation} \label{nonlinFeq}
\partial \left(\frac{\partial {\hat F} }{1+ {\hat F}} \right) + \left(\frac{\partial {\hat F}}{1+ {\hat F}} \right)^2 +\kappa^2  {\hat F} =0~.
 \end{equation}

We have computed the explicit form of this derived field equation for ${\hat F}_{(ij)}$ up to the quadratic order in  ${\hat F}_{(ij)}$. 
Let us display our results when using the variable $u_{ij}=u_{(ij)}$, defined as %
\begin{align}
    u_{ij}&\equiv \frac{1}{\kappa^2} \left( F_{(ij)}-\frac{1}{6}\eta_{ij}F\right).
\end{align}
For brevity, we do not henceforth explicitly indicate that 
$u_{ij}$ is a symmetric tensor. We have followed here Refs. \cite{Nikiforova:2009qr,Deffayet:2011uk} in defining $u_{ij}$ so that it satisfies, at the linear level, the usual Fierz-Pauli equation. We
shall show below how to derive Eq. \eqref{nonlinFeq}  from an action.

\subsection{Linear order equation for $u_{ij}$}

The first step to obtain the linear equation for $u_{ij}$ is to invert the (linear) relation Eq.~\eqref{eqn:Hijk}  between  $H_{ijk}$
and $K_{ijk} = A_{ijk}$. An easy computation yields 
\begin{align}
\label{eqn:eqofA}
    A_{ijk}=&\frac{1}{2 c_{F}}\left[H_{ijk}-H_{jki}-H_{kij}-\frac{1}{2}\eta_{jk}\left(2H\indices{_{il}^l}-H\indices{^l_{li}}\right)\right.\nonumber \\
    &\left.+\frac{1}{2}\eta_{ik}\left(2H\indices{_{jl}^l}-H\indices{^l_{lj}}\right)\right]~.
\end{align}
Inserting Eq.~\eqref{eqn:EOMAnew} in this expression, then yields the 
explicit form of the equation sketched in Eq.~\eqref{AvsF0} above. Proceeding as just indicated  and working at the linear level in $u_{ij}$ yields the linear equation
\begin{equation}
 \mathcal{U}_{ij} =0, 
\end{equation}
where 
\begin{align}
\label{eqn:EOMofU}
    \mathcal{U}_{ij}&\equiv\partial_k\partial^k u_{ij}+\partial_i\partial_j u-\partial_i\partial_ku\indices{_j^k}-\partial_j\partial_k u\indices{_i^k}\nonumber\\
    &+\eta_{ij}\left(\partial_k\partial_l u^{kl}-\partial_k\partial^k u\right)-\kappa^2\left(u_{ij}-\eta_{ij}u\right).
\end{align}
The equation $\mathcal{U}_{ij}=0$ is the usual form of the (linear) Fierz-Pauli equation \cite{Fierz:1939ix}. Let us recall that one way to prove that this equation describes the five dof of a massive spin-2 field is to take 
the divergence and the trace of $\mathcal{U}_{ij}=0$. These yield the constraints
\begin{align} 
\label{lincons1a}
    \partial^j \mathcal{U}_{ij}&\equiv\kappa^2 ( \partial_i u-\partial^j u_{ij})=0~,\\
    \label{lincons1b}
  \eta^{ij} \mathcal{U}_{ij} + \frac{2}{\kappa^2} \partial^{ij} \mathcal{U}_{ij} &\equiv  -3 \kappa^2 u =0~.
\end{align}
These $4+1$ constraints (last equations on the right) involve only $u$ and $\partial u$ and are equivalent to the following five constraints,
\begin{align} 
\label{lincons2}
    \partial^j u_{ij} =0~,\hspace{0.8cm} u=0~,
\end{align}
which ensure the correct number (five) of dof  at the linear level.

The main aim of our present study is to investigate whether this number of dof is preserved at the nonlinear level.

\subsection{Next-to-linear order}

As a first attempt to investigate the healthiness of the nonlinear massive-gravity theory described by the sketchy equation Eq.~\eqref{nonlinFeq} we explicitly computed the terms quadratic in $F_{(ij)}$ (i.e in $u_{ij}$)
in Eq.~\eqref{nonlinFeq} by following the procedure explained above.
We obtained
\begin{widetext}
    \begin{align} \label{U=Uquad}
    \partial_k\partial^k u_{ij}&+\partial_i\partial_j u-\partial_i\partial_ku\indices{_j^k}-\partial_j\partial_k u\indices{_i^k}+\eta_{ij}\left(\partial_k\partial_l u^{kl}-\partial_k\partial^k u\right)-\kappa^2\left(u_{ij}-\eta_{ij}u\right)=\mathcal{U}^{\rm quad}_{ij}(u)+ \mathcal{O}(u^3)~,
   \end{align}
   with an effective quadratically nonlinear source given by
   \begin{align}
    \mathcal{U}^{\rm quad}_{ij}(u)&=\partial_i u_{kl}\partial_j u^{kl}-\partial_i u~\partial_j u + u^{kl}\partial_i\partial_j u_{kl}-u~\partial_i \partial_j u+\partial^k u_{ik}\partial_ju +\partial^k u_{jk}\partial_i u \nonumber\\
    &+\frac{1}{2}\left(u_{jk}\partial^k\partial_iu+u_{ik}\partial^k\partial_ju\right)-\partial_k u\partial^k u_{ij}-\partial^k u_{ik}\partial^l u_{jl}+\frac{1}{2}\left(\partial_iu_{jk}\partial_lu^{kl}+\partial_ju_{ik}\partial_l u^{kl}-u^{kl}\partial_l\partial_iu_{jk}-u^{kl}\partial_l\partial_j u_{ik}\right)\nonumber \\
    &+u\partial^l\partial_i u_{jl}+u\partial^l\partial_j u_{il}-u_{jk}\partial_l\partial_i u^{kl}-u_{ik}\partial_l\partial_j u^{kl}+u_{ij}\partial_l\partial_k u^{lk}-u\partial_l\partial^lu_{ij}+u\indices{_j^k}\partial_l\partial^lu_{ik}+u\indices{_i^k}\partial_l\partial^lu_{jk}-u_{ij}\partial_l\partial^lu\nonumber\\
    &+\partial^ku_{jl}\partial^lu_{ik}+\partial^ku_{jl}\partial^ku_{il}-\frac{1}{2}\left(u_{jk}\partial^l\partial^k u_{il}+u_{ik}\partial^l\partial^k u_{jl}\right)-\frac{3}{2}\left(\partial_ju^{kl}\partial_lu_{ik}+\partial_{i}u^{kl}\partial_lu_{jk}\right)
    \nonumber\\
    &-\eta_{ij}\left[\frac{1}{2}\left(\partial^n u^{kl}\partial_n u_{kl}+\partial_ku^{kl}\partial^nu_{nl}-\partial^l u\partial_lu-\partial_l u_{kn}\partial^n u^{kl}\right)-u\partial_k\partial^k u-u^{kl}\partial^n\partial_lu_{nk}+u\partial_n\partial_l u^{nl}+u^{kl}\partial_n\partial^nu_{kl}\right]~.
    \end{align} 
\end{widetext}

If the nonlinear massive gravity theory described
by Eq.~\eqref{U=Uquad} had the same number of dof as its linear counterpart, one would expect 
Eq.~\eqref{U=Uquad} to give rise again to five quadratically deformed first-order constraints of the type
\begin{align}
    \partial^j u_{ij} =\mathcal{O}_i^{\rm quad}(u, \partial u)~,\hspace{0.8cm} u=\mathcal{O}^{\rm quad}(u, \partial u)~.
\end{align}
One would expect that such quadratically deformed first-order constraints can be revealed
by replacing $~\mathcal{U}_{ij}(u)$ on the left-hand-sides of Eqs.~\eqref{lincons1a} and \eqref{lincons1b} by its quadratically deformed avatar $\mathcal{U}_{ij}(u) -\mathcal{U}^{\rm quad}_{ij}(u)$. If that were the case, the quadratic quantities
\begin{align} 
\label{lincons3}
     C^{\rm quad}_i &\equiv \partial^j \mathcal{U}^{\rm quad}_{ij},\\
 C^{\rm quad} &\equiv \eta^{ij} \mathcal{U}^{\rm quad}_{ij} + \frac{2}{\kappa^2} \partial^{ij} \mathcal{U}^{\rm quad}_{ij} ,
\end{align}
would be expected to vanish when working
on linear shell ($\mathcal{U}_{ij}(u)=0$),
and imposing the
linear-level constraints given in Eq.~\eqref{lincons2}.

However, we found that the so-defined quantities
$ C^{\rm quad}_i,  C^{\rm quad}$ do not vanish when imposing Eq.~\eqref{lincons2}, but, actually, involve up to three derivatives of $u$.

This result suggests that the (second-order) nonlinear massive gravity equations satisfied by $u_{ij}$ propagate more than five dof, or exhibit some worse inconsistency.
However, this reasoning does not allow one to understand the real number of dof of the exact field equations for $u_{ij}$ whose structure was
indicated in Eq.~\eqref{nonlinFeq}. In addition, the reasoning leading to the derivation of Eq.~\eqref{nonlinFeq} did not allow us to prove that this second-order field equation for $F_{(ij)}$ (or $u_{ij}$) could be derived from an action.
In the following sections, we shall therefore take a different tack (directly based on action principles).

\section{Action formulations of torsion bigravity}
\label{Sec:ActionformTG}

In the decoupling limit $c_R\rightarrow \infty$ (with the Eq.~\eqref{gexp} assumption) considered here, the torsion bigravity theory is originally defined by a {\it second-order} action depending only on the connection $A_3 = A_{ijk}= A_{[ij]k}$, say [see, Eq.~\eqref{eqn:Lagdec}], 
\begin{align}
\label{Lagnew1}
    \mathcal{L}_{TBG} [A_3] =&~c_{F^2} \left(F_{(ij)}(A, \partial A)F^{(ij)}(A, \partial A)\right.\nonumber\\
    &\left.-\frac{1}{3}F^2(A, \partial A)\right)+c_{F} F(A, \partial A)~.
\end{align}
Let us now show that torsion bigravity also admits two other useful action formulations:
(1) a {\it first-order} Lagrangian  $ \mathcal{L}_{1}(b_2, A_3)$ involving both  
$A_3$ and a symmetric two-tensor $b_2=b_{(ij)}$
(equivalent to $F_{(ij)}$ or  $u_{ij}$); and (2) 
a {\it second-order} Lagrangian  $  \mathcal{L}_{2}(b_2)$ involving only $b_2=b_{(ij)}$ (and leading to the second-order equation of motion given in Eq.~\eqref{nonlinFeq}).

These alternative action formulations are derived from the original action Eq.~\eqref{Lagnew1} by introducing both a Lagrange multiplier $b_{ij}=b_{(ij)}$ and an auxiliary
independent symmetric  field $\mathcal{F}_{ij}$.

 Indeed, we can construct a Lagrangian depending on the three fields $[b_2, A_3, \mathcal{F}_2]$ such that $b_{ij}$ acts as a Lagrange multiplier imposing the constraint $\mathcal{F}_{ij} =F_{ij}(A)$.
A Lagrangian satisfying these requirements is
\begin{align}
\label{eqn:Slagxier}
    \mathcal{L}_{0}[b_2, A_3, \mathcal{F}_2] =&  c_{b} \, b^{ij}\left(\mathcal{F}_{ij}-F_{ij}(A, \partial A)\right)+c_F \mathcal{F} \nonumber \\
    &+c_{F^2} \left(\mathcal{F}_{ij}\mathcal{F}^{ij}-\frac{1}{3}\mathcal{F}^2\right)~,
\end{align}
where $c_b$ is a constant (to be chosen below), and $\mathcal{F}\equiv\mathcal{F}_{ij}\eta^{ij}$. The variation of $\mathcal{L}_0$ with respect to $b_{ij}$ gives the constraint $\mathcal{F}_{ij}=F_{ij}(A)$. We can use this constraint to eliminate both $b_{ij}$ and $\mathcal{F}_{ij}$ from the Lagrangian. This then reduces $\mathcal{L}_0[b_2, A_3, \mathcal{F}_2]$ to the original Lagrangian $\mathcal{L}_{TBG[A_3]}$.

On the other hand, as the action $\mathcal{L}_0[b_2, A_3, \mathcal{F}_2]$,
Eq.~\eqref{eqn:Slagxier}, is
quadratic in the independent auxiliary field
$\mathcal{F}_{ij}$,  we can instead ``integrate out"  $\mathcal{F}_{ij}$ by solving its equation
of motion $ \frac{\delta \mathcal{L}_0}{\delta \mathcal{F}_{ij}}=0$. The latter equation of motion reads,
\begin{align}
\label{eqn:Feliminate}
    \frac{\delta \mathcal{L}_0}{\delta \mathcal{F}_{ij}}= c_b b_{ij} +c_F \eta_{ij} + 2 c_{F^2}\left(\mathcal{F}_{ij}-\frac{1}{3}\eta_{ij}\mathcal{F}\right) =0~.
\end{align}
It is convenient to choose
\begin{equation}
 c_b = -c_F ,   
\end{equation}
to ensure that $b_{ij}\rightarrow \eta_{ij}$ when $\mathcal{F}_{ij}\rightarrow 0$.  Doing this, the field equation for $\mathcal{F}_{ij}$ becomes (recalling $ \mathcal{F} \equiv \eta^{ij} \mathcal{F}_{ij}$)
\begin{align}
    \mathcal{F}_{ij}-\frac{1}{3}\eta_{ij}\mathcal{F}=\frac{\kappa^2}{2}( b_{ij} - \eta_{ij})~.
\end{align}
Defining for notational convenience
\begin{equation}
 \bar{b}_{ij} \equiv b_{ij}-\eta_{ij}~,
\end{equation}
the above equation yields (with $\bar{b} \equiv \eta^{ij} \bar{b}_{ij}$)
\begin{align}
\label{eqn:Fel1}
    \mathcal{F}_{ij}=\frac{\kappa^2}{2}\left(\bar{b}_{ij}-\eta_{ij}\bar{b}\right)~,~~~~ \mathcal{F} = -\frac{3\kappa^2}{2}\bar{b} \, .
\end{align}
Inserting Eq.~\eqref{eqn:Fel1} in Eq.~\eqref{eqn:Slagxier}, we then obtain the following first-order action involving $b_2$ and $A_3$,
\begin{align}
\label{eqn:SLag2xier}
    \mathcal{L}_1[b_2, A_3]=c_F b^{ij}F_{ij}(A)-\frac{c_F\kappa^2}{4}\left(\bar{b}_{ij}\bar{b}^{ij}-\bar{b}^2\right)~.
\end{align}
This first-order action will be used below to investigate the dof of torsion bigravity in the decoupling limit.

The Lagrangian given in Eq.~\eqref{eqn:SLag2xier} is linear in $F_2(A) = \partial A_3 + A_3 A_3$, and therefore {\it quadratic} in $A_3$. We can therefore ``integrate out" $A_3$ from the Lagrangian $\mathcal{L}_1[b_2, A_3]$, Eq.~\eqref{eqn:SLag2xier}, by solving its equation of motion $\frac{\delta \mathcal{L}_1}{\delta A_{3}}=0$.

Using the explicit
expression of $F_2(A) = \partial A_3 + A_3 A_3$, namely
\begin{align}
    F_{ij}(A) =& \frac{1}{2}\left(\partial_{j}A\indices{_{il}^l}-\partial^{l}A_{ilj}+A_{imj}A\indices{^{ml}_l}-A\indices{_i^{ml}}A_{mlj} \right.\nonumber\\
    &\left. +~{i\leftrightarrow j}\right),
\end{align}
we find that the field equation for $A_3$ reads
\begin{align}
    \frac{\delta \mathcal{L}_1}{\delta A_{ijk}} =& \frac{1}{2} c_F\left[ \beta_{ijk}
    - (b_2 * A_3)_{ijk}\right]
     =0~,
\end{align}
where we define the following
rank-3 tensors having the same symmetries as $A_{ijk}=A_{[ij]k}$,
\begin{align} \label{defbeta}
    \beta_{ijk} \equiv \partial_{i}b_{jk}-\eta_{ik}\partial^lb_{jl}- (i\leftrightarrow j)~,
\end{align}
and
\begin{align}
   (b_2 * A_3)_{ijk}  \equiv& 
   b_{ik}A\indices{_{jl}^l}-b_{il}A\indices{_{jk}^l}+b_{kl}A\indices{_i^l_j}+\eta_{ik}b^{lm}A_{jlm}
\nonumber \\
&- (i\leftrightarrow j)~.
\end{align}

The field equation of $A_3$ is then linear (and algebraic) in $A_3$ and has the structure 
\begin{equation} \label{eomAA}
    (b_2 * A_3)_{ijk} =   \beta_{ijk}\,,
\end{equation}
where $\beta_{ijk}$ is the (curl-like) linear
combination of the first derivatives of $b_{ij}$ defined in Eq.~\eqref{defbeta}. 

The structure of the field equation, Eq.~\eqref{eomAA}, is reminiscent of the Lyapunov
matrix equation
\begin{align} \label{lyapunov}
B X + X B = C~,
\end{align}
where $X$ is an unknown matrix (playing the role of $A_3$) while the matrix $B$ plays the role of
$b_2$ and the right-hand-side matrix $C$ plays the role of $\beta_3 \sim \partial_1 b_2$. The solution of this Lyapunov toy model is easily obtained if the matrix $B$ is diagonalizable.
Indeed, when working in a frame where $B_{ij} = b_i \delta_{ij}$, the equation for $ X_{ij}$
reads $(b_i + b_j)X_{ij} = C_{ij}$, which is solved
as 
\begin{align}
    X_{ij} = \frac{C_{ij}}{(b_i + b_j)}~,
\end{align}
if there are no vanishing denominators.

We can morally think of the matrix denominators
$(b_i + b_j)^{-1}$ as defining some new ``inverse" $B^{-1}$ of the matrix $B$.
In fact, if the matrix $B$ is close to the
unit matrix, say $B_{ij}= \delta_{ij} + {\bar b}_{ij}$ (which is similar to our situation if we work in the Euclidean signature), we can expand this special-frame solution
in powers of the eigenvalues ${\bar b}_i$ 
of ${\bar b}_{ij}$ and thereby easily rewrite
 the general-frame solution $ X_{ij}$ in terms
 of a series of terms made of various contractions of $C_{ij}$ with several factors
 involving ${\bar b}_{ij}$. Suppressing indices,
 signs and factors, we have the structure
 \begin{equation}
 X_2 \sim C_2 + {\bar b}_2 * C_2 + {\bar b}_2*{\bar b}_2 * C_2 + \cdots   ~. 
 \end{equation}

It is easily seen that the obtention of such a perturbative solution to the Lyapunov equation 
Eq.~\eqref{lyapunov} can be extended to the linear $A_3$ equation Eq.~\eqref{eomAA} when
expanding its solution in powers of ${\bar b}_{ij}= b_{ij} - \eta_{ij}$. Namely, the solution
for $A_3$ will have the perturbative structure
(valid for ${\bar b}_{ij} < O(1)$)
 \begin{equation}
 A_3 \sim \beta_3 + {\bar b}_2 * \beta_3 + {\bar b}_2*{\bar b}_2 * \beta_3 + \cdots    ~.
 \end{equation}

Actually, the successive terms in this expansion can be straightforwardly derived by solving the explicit form of the $A_3$ equation, namely,
\begin{align}
\label{eqn:NTbA}
     \beta_{ijk}= \left[b_{ik}A\indices{_{jl}^l}-b_{il}A\indices{_{jk}^l}+b_{kl}A\indices{_i^l_j}+\eta_{ik}b^{lm}A_{jlm}- (i\leftrightarrow j)\right],
\end{align}
in an expansion in powers of ${\bar b}_{ij}= b_{ij} - \eta_{ij}$.

The first term in this expansion is obtained by
replacing $b_{ij} \to \eta_{ij}$ on the right-handside of Eq.~\eqref{eqn:NTbA}. This yields
\begin{align}
    A^{LO}_{ijk}=\frac{1}{2} \left( \beta_{ijk}-\beta_{kij}+\beta_{jki} \right)~,
\end{align}
which is actually of order $ A^{LO} \sim \partial {\bar b}$. We also explicitly computed
the next term in the expansion
\begin{equation}
    A_3[{\bar b}_2, \partial {\bar b}_2] =  A^{LO}_3 +  A^{NLO}_3 + \cdots\sim \partial {\bar b}_2 + {\bar b}_2 \partial {\bar b}_2+ \cdots~.
\end{equation}
We can morally think that this solution is the massive-gravity analog of the solution expressing the Levi-Civita connection in terms of the (inverse of the) metric $g_2$ and of its first derivatives $\partial g_2$. We can also formally write its structure as
\begin{equation} \label{A3sol}
    A_3[{\bar b}_2, \partial {\bar b}_2] \sim (b_2)^{-1}  \beta_3  \sim (b_2)^{-1} \partial_1  b_2~.
\end{equation}

Integrating out $A_3$ from the action 
$\mathcal{L}_1[b_2, A_3]$, Eq.~\eqref{eqn:SLag2xier}, simply means replacing
$A_3$ in $\mathcal{L}_1[b_2, A_3]$ 
(after an integration by parts to shuffle the derivatives from $A_3$ to $b_2$) by the
solution Eq.~\eqref{A3sol}. This yields a
second-order Lagrangian for torsion bigravity depending only on $b_2$, and having the structure
\begin{equation} 
\label{eq:L2b}
\frac{\mathcal{L}_2 [b]}{c_F} \sim  \partial  b *b^{-1}* \partial  b + b * b^{-1}* \partial  b* b^{-1}* \partial  b + \kappa^2 {\bar b}^2 \,.
\end{equation}
The kinetic terms in this Lagrangian are defined by a power series expansion in ${\bar b}_{ij}$, while the mass terms are quadratic in  ${\bar b}_{ij}$ and exactly given by the last term in
Eq.~\eqref{eqn:SLag2xier}. Namely, 
\begin{equation*}
\frac{\mathcal{L}^{\rm mass \; term}_2 [b]}{c_F} =  -\frac{1}{4} \kappa^2  \left(\bar{b}_{ij}\bar{b}^{ij} - \bar{b}^2 \right).
\end{equation*}
Note in passing that the present reasoning shows that the nonlinear equation \eqref{nonlinFeq} can indeed be derived from an action, namely Eq.~\eqref{eq:L2b}. 

Forgetting that it was obtained by decoupling a dynamical torsionfull connection from an associated dynamical metric, the second-order action, Eq.~\eqref{eq:L2b}, considered as a Poincar\'e-invariant action in Minkowski spacetime for $b_{ij}$ defines a 
specific nonlinear massive gravity theory, which is quite different from the usually considered massive gravity theories. Indeed, the latter \cite{Damour:2002ws,deRham:2014zqa} define the kinetic term of $b_{ij}$ by means of the Einstein-Hilbert action, and then look for a nonlinear deformation of the Fierz-Pauli mass term having good properties. By contrast the Einstein-Cartan origin of the dynamics of 
\begin{equation} \label{bvsF}
b_{ij}= \eta_{ij} + \frac{2}{\kappa^2} ({\mathcal F}_{ij} -\frac13 {\mathcal F} \eta_{ij}) \,,
\end{equation}
 has led to the nonlinear action  $\mathcal{L}_2 [b]$,  Eq.~\eqref{eq:L2b}, which features a  mass term of the usual quadratic Fierz-Pauli form, but with specific, non-Einsteinian kinetic terms.

We have explicitly computed the second-order action $\frac{\mathcal{L}_2 [b]}{c_F}$ up to cubic order in ${\bar b}_{ij}$, i.e.
\begin{equation*} \label{L2bexp}
\frac{\mathcal{L}_2 [b]}{c_F} =  \frac{\mathcal{L}^{LO}_2 [b]}{c_F}+ \frac{\mathcal{L}^{NLO}_2 [b]}{c_F}+ \cdots
\end{equation*}
The leading-order action (quadratic in ${\bar b}_{ij}$) is

\begin{align}
\frac{\mathcal{L}^{LO}_2 [b]}{c_F} =&\frac{1}{8}\partial_k\bar{b}\indices{_i^i}\partial^k\bar{b}\indices{_j^j}+\frac{1}{4}\partial_j\bar{b}^{jk}\partial^i\bar{b}_{ik}+\frac{1}{4}\partial_k\bar{b}_{ij}\partial_i\bar{b}^{jk}\nonumber\\
&-\frac{1}{4}\partial_i\bar{b}^{jk}\partial^i\bar{b}_{jk} -\frac{1}{4} \kappa^2  \left(\bar{b}_{ij}\bar{b}^{ij} - \bar{b}\indices{_i^i}\bar{b}\indices{_j^j}\right)~,
\end{align}
while the next to leading-order one (cubic in ${\bar b}_{ij}$) reads
\begin{widetext}

\begin{align}
\frac{\mathcal{L}^{NLO}_2[b]}{c_F} = & \frac{1}{8}\bar{b}_{ij}\left(\beta\indices{^j_{kl}}\beta^{kli}-\frac{1}{2}\beta\indices{_{kl}^j}\beta^{kli}-\beta^{jki}\beta\indices{_{kl}^l}\right)+\frac{1}{8}\bb_{ij}\beta\indices{_{kl}^l}\partial^i\bar{b}^{jk}+\frac{1}{8}\bb^{ij}\beta\indices{_{jk}^k}\partial^l\bb_{il}-\frac{1}{8}\bb\indices{_i^i}\beta\indices{_{jk}^k}\partial_l\bb^{lj}
\nonumber\\
&+\frac{1}{8}\bb\indices{_{ij}}\beta\indices{^{ikj}}\partial_k\bb\indices{_l^l}-\frac{1}{4}\bb\indices{^{ij}}\beta\indices{_{ikj}}\partial_l\bb\indices{^{lk}}-\frac{1}{2}\bb_{ij}\beta\indices{_l^{ki}}\partial_k\bb^{lj}-\frac{1}{8}\bb_{ij}\beta\indices{^{kl}_l}\partial_k\bb^{ij}~.
\end{align}    
\end{widetext}

We have checked that the equations of motion for ${\bar b}_{ij}$
derived 
from the above Lagrangian agree with the equations of motion Eq.~\eqref{nonlinFeq}. To check the agreement, one must take into account the relation between ${\bar b}_{ij}$ and $F_{ij} = \mathcal F_{ij}$ given by Eq.~\eqref{eqn:Fel1}.
In particular, the latter relation implies that ${\bar b}_{ij}$ is related to the
Fierz-Pauli usual variable $u_{ij}$ via
\begin{align}
\bar{b}_{ij} = u_{ij}-\frac{1}{2}\eta_{ij} u~.
\end{align}
Note in passing the curious fact that the Fierz-Pauli mass term has the same expression in terms of ${\bar b}_{ij}$ and in terms of $u_{ij}$, because
\begin{equation}
   \bar{b}_{ij}\bar{b}^{ij} - \bar{b}^2  \equiv u_{ij}u ^{ij} - u^2  \,.
\end{equation}

\section{Structure of the First-order Action}
\label{Sec:Structureaction}

In this section, we finally tackle in a precise manner the issue of the number of dof of torsion bigravity at the nonlinear level by using
 the first-order  Lagrangian  $\mathcal{L}_1[b,A]$, Eq.~\eqref{eqn:SLag2xier}, namely: 
\begin{align}
\label{eqn:first-orderL}
\frac{\mathcal{L}_1[b_2,A_3]}{c_F}=  b^{ij}F_{ij}(A)-\frac{\kappa^2}{4}\left(\bar{b}_{ij}\bar{b}^{ij}-\bar{b}^2\right)~.
\end{align}
In order to reach clear conclusions, we shall focus here on the particular case where (after 
having performed the field transformations leading to Eq.~\eqref{eqn:first-orderL}) all the variables entering the action $\mathcal{L}_1[b_2,A_3]$ are homogeneous in space and only depend on time, $t= x^0$.

Remembering that $F[A]= \partial A + A A$, we see from Eq.~\eqref{eqn:first-orderL}, that, if we symbolically denote the variables entering the first-order action as $ q = A_3$ and $p={\bar b}_{ij}$ we have (when all variables depend only on time, and after discarding the total-derivative term $\eta^{ij} F^{\rm lin}_{ij}[A]\sim \dot A $)
\begin{equation}
\frac{\mathcal{L}_1[p,q]}{c_F}= p \dot q - p q q - \kappa^2 p^2\,.
\end{equation}
In other words, we have a Hamiltonian action
$ p \dot q - H(p,q)$ with a cubic Hamiltonian
$H(p,q)=p q q + \kappa^2 p^2 $.

This would be a trivial problem if this action did not involve many different constraints. Our task will be to delineate the precise constrained structure of the first-order action Eq.~\eqref{eqn:first-orderL}.

In the following, we find it convenient to work with the metriclike Fierz-Pauli variable
$u_{ij}$, instead of $\bar{b}_{ij}$. To avoid
confusion with the various notations used above, we shall actually henceforth denote this 
Fierz-Pauli variable as $h_{ij}$. In other words,
we henceforth parametrize $b_{ij}$ as
\begin{align}
    b_{ij}& \equiv \eta_{ij}+\bb_{ij}\nonumber\\
    &\equiv \eta_{ij}+h_{ij}-\frac{1}{2}\eta_{ij}h\indices{_k^k}~.
\end{align}
Let us first consider the kinetic $p \dot q$ component of the first-order Lagrangian $\frac{\mathcal{L}_1}{c_F}$, namely
\begin{align}
\label{eqn:pdq}
    p \dot q =b_{00}\tilde{F}_{00}-b_{0a}\tilde{F}_{0a}+b_{ab}\tilde{F}_{ab}\,,
\end{align}
where $\tilde{F}$ denotes the $ \dot q \sim \dot A$ part of the Ricci tensor $F[A]= \partial A + A A$. We find (using our $(-,+,+,+)$ signature)
\begin{align}
    \tilde{F}_{00} &= \partial_0 A\indices{_{0a}^a}~,\\
    \tilde{F}_{0a} &= \frac{1}{2}\partial_0 A\indices{_{ac}^c}~,\\
    \tilde{F}_{ab} &= \frac{1}{2}\left(\partial_0A_{a0b}+\partial_0A_{b0a}\right).
\end{align}
Here and henceforth, the indices $a,b,c$ denote spatial indices, $a,b,c = 1,2,3$ (moved by the 
3-dimensional Euclidean metric $\delta_{ab}$), in contrast to the indices $i,j,k,\cdots$ used above to denote spacetime frame indices $i,j,k,\cdots= 0,1,2,3$ (moved by the 
4-dimensional Minkowski metric $\eta_{ij}$). [Remember that in our decoupling limit, the frame indices can be identified with Lorentz-coordinate indices $\mu = i$.]

Substituting the above equations in Eq.~\eqref{eqn:pdq}, we obtain,
\begin{align}
\label{eqn:L2pdq}
    p \dot q &= \frac{1}{2}h_{00}\left(\tilde{F}_{00}+\tilde{F}\indices{_a^a}\right)+\frac{1}{2}\bar{h}\left(\tilde{F}_{00}-\tilde{F}\indices{_a^a}\right)-h_{0a}\tilde{F}\indices{_0^a}\nonumber\\
    &\quad+\tilde{F}_{ab}h^{ab}~\nonumber\\
    &=\bar{h}\partial_0A\indices{_{0a}^a}-\frac{1}{2}h_{0a}\partial_0A\indices{_{ab}^b}-\frac{1}{2}h_{ab}\left(\partial_0A_{0ab}+\partial_0A_{0ba}\right)~\nonumber\\
    &=-\frac{1}{2}h_{0a}\partial_0A\indices{_{ab}^b}-h_{ab}\left(\partial_0A_{0(ab)}-\delta^{ab}\partial_0A\indices{_{0c}^c}\right)~,
\end{align}
where the bar in $\bar{h}$ denotes a trace only on the space indices: $\bar{h} \equiv \delta_{ab}h^{ab}$. Here we have also used 
\begin{align}
\label{eqn:pdqpqq}
    \partial_0A_{0(ab)} = \frac{1}{2}\left(\partial_0A_{0ab} +\partial_0A_{0ba} \right)~.
\end{align}
The above equation, Eq.~\eqref{eqn:L2pdq}, shows that the ``momentum" $p_{00}=h_{00}$ has disappeared in the $p \dot q$ kinetic terms. The
momentum $p_{00}=h_{00}$  can then only appear in the $pqq \sim hAA$ terms, or the $p^2$ ones. Evidently, it can only appear linearly in the $pqq \sim hAA$ terms. Concerning the  $p^2$ ones, it is well known that the (quadratic) Fierz-Pauli mass term is such that $p_{00}=h_{00}$ only appears linearly. Indeed, from Eq.~\eqref{eqn:first-orderL},
\begin{align}
\label{eqn:p2}
   \kappa^2 p^2 &= \frac{\kappa^2}{4}\left(\bar{b}_{ij}\bar{b}^{ij}-\bar{b}^2\right)\nonumber\\
   &=\frac{\kappa^2}{4}\left(h_{ij} h^{ij}-(\eta^{ij} h_{ij})^2\right)\nonumber\\
    &=\frac{\kappa^2}{4}\left(h_{ab}h^{ab}-\bar{h}^2-2h_{0a}^2+2h_{00}\bar{h}\right)~.
\end{align}
Hence, we conclude that in torsion bigravity $h_{00}$ is an exact Lagrange multiplier whose equation of motion yields a scalar constraint equation. This situation is quite different from the usual massive gravity theories where the lapse appears linearly in the $ p \dot q$ terms but nonlinearly in the mass term, and where the existence of a scalar constraint in ghost-free theories is rather hidden\cite{deRham:2010kj,deRham:2011rn,Hassan:2011ea,Hassan:2011hr}. 

Let us now continue our exploration of the constrained nature of the first-order action, 
Eq.~\eqref{eqn:first-orderL}. 
 Eq.~\eqref{eqn:L2pdq} shows that amongst a priori possible twenty-four kinetic
 terms $\dot q = \partial_0A_{ijk}$, i.e. $\partial_0A_{0(ab)}\, ,\partial_0A_{0[ab]}\, ,\partial_0A_{a00}\, ,\partial_0A_{[ab]0}\, ,\partial_0A_{[ab]c}$, only nine combinations, namely $\partial_0A_{0(ab)}\,$, and $\partial_0A\indices{_{ab}^{b}}$, enter the first-order Lagrangian $\mathcal{L}_1$. Let us define as follows the corresponding nine $q$ variables entering $ p \dot q$ terms,
\begin{align} \label{defq}
    q_a&\equiv-A\indices{_{ab}^b}~,\\ \nonumber
    q_{(ab)}&\equiv-A_{0(ab)}+\delta_{ab}A\indices{_{0c}^c}~.
\end{align}
Besides the nine ``active" $q$'s, Eq.~\eqref{defq}, entering the kinetic terms, the first-order action contains fifteen remaining $q$-type variables which enter the Hamiltonian
$H(p,q)$ without participating in the kinetic terms. These remaining fifteen $q$-type variables happen to include two different types of variables. Six of them turn out to appear
linearly in the action (and therefore to be Lagrange multipliers), while nine of them appear quadratically in the action. 

The six Lagrange-multiplier $q$ variables (decorated with a superscript $L$, for Lagrange) are
\begin{align}
    {Q}^L_a&=A_{a00}~,\\
    Q^L_{[ab]}&= A_{[ab]0}=\frac{1}{2}\left(A_{ab0}-A_{ba0}\right)~.
    \end{align}
Because of the antisymmetry over $ab$ in $ Q^L_{[ab]}$ there are only three different two-index Lagrange multipliers.

The remaining nine $q$-type variables entering
the action quadratically are
\begin{align}
    \check{Q}_{[ab]}&= A_{0[ab]}=\frac{1}{2}\left(A_{0ab}-A_{0ba}\right)~,\\
    \check{Q}_{abc}&=A_{[ab]c}-\frac{1}{2}\left(A\indices{_{ad}^d}\delta_{bc}-A\indices{_{bd}^d}\delta_{ac}\right)~,
\end{align}
where $\check{Q}_{abc}$ is anti-symmetric in $(a\leftrightarrow b)$ and traceless. 
The $ab$ anti-symmetry of $\check{Q}_{[ab]}$ means that they contain only three independent components.
The algebraic constraints on $\check{Q}_{abc}$
mean that these contain only six independent components. To derive the equations of motion of $\check{Q}_{abc}$, we will express it in terms of its symmetric dual tensor $\check{Q}^*_{(ab)}$ as
\begin{align}
\label{eqn:dual}
    \check{Q}_{abc} = \epsilon_{abd}\check{Q}^*_{(cd)}~.
\end{align}
At this stage, the first-order Lagrangian  can  be written as
\begin{align}
    \frac{\mathcal{L}_1}{c_F} = h_{ab}\dot{q}^{ab}+\frac{1}{2}h_{0a}\dot{q}_a-\mathcal{H}(h_{ab},h_{00},h_{0a},q_{ab},q_{a},Q^L, \check{Q})~,
\end{align}
where we recall that $\mathcal{H} = pqq+p^2 = hAA +hh$. 

As announced, this first-order
action involves seven Lagrange multipliers,
namely $h_{00}$, ${Q}^L_a$ and ${Q}^L_{[ab]}$.
This means that the Hamiltonian part
of the action can be written as

\begin{align} \label{Hexact}
-\mathcal{H} &= h_{00} C[{\rm vars}] +  {Q}^L_a C_a[{\rm vars}] +\frac12 {Q}^L_{[ab]} C_{[ab]}[{\rm vars}] \nonumber\\ 
 &-\mathcal{H}^{\rm remain}[{\rm vars}]~,
\end{align}
where $[{\rm vars}]$ denote the following (initial) list of variables
\begin{equation}
    [{\rm vars}]= [h_{ab},q_{ab},h_{0a},q_{a}, \check{Q}_{[ab]}, \check{Q}_{abc}]\,.
\end{equation}
Note that the seven Lagrange multipliers imply
seven corresponding (first-class) constraints, namely
\begin{align} \label{sevenLcons}
& C[{\rm vars}] =0, \\ \nonumber
& C_a[{\rm vars}] =0, \\ \nonumber
& C_{[ab]}[{\rm vars}] =0.
\end{align}
However, the impact of these seven constraints depend on their precise form. 
In addition, the fact that the six variables $\check{Q}_{abc}$ enter
only algebraically (and quadratically) in the action
means that they will also give rise to six (second-class) constraints, namely
\begin{equation}
   \frac{\partial \mathcal H}{\partial \check{Q}^*_{(ab)}} =0\,.
\end{equation}
The latter additional constraints are best taken into account after having dealt with the
seven Lagrange-multiplier constraints, Eq.~\eqref{sevenLcons}.

To clarify the difference between the linear and the nonlinear dynamics of torsion bigravity, we will consider first the leading-order (linear)
dynamics.

\section{Torsion bigravity at the linear level}
\label{Sec:TGLinear}
At the leading-order (linear equations of motion), 
the Hamiltonian has the general form \eqref{Hexact} with
\begin{align}
 C^{LO} &=-\frac{\kappa^2}{2} \bar h~, \\
 C^{LO}_a &=2 q_a ~, \\ 
 C^{LO}_{[ab]} &=4 \check{Q}_{[ab]}~,
\end{align}
and
\begin{align}
  -\mathcal{H}^{\rm remain} &= -\frac{1}{2}q_a^2+\frac{1}{2}\left(q\indices{_a^a}\right)^2-q_{ab}q^{ab}+\check{Q}_{ab}\check{Q}^{ab}\nonumber\\
    &+\check{Q}_{abc}\check{Q}^{cba}-\frac{\kappa^2}{4}\left(h_{ab}h^{ab}-\bar{h}^2-2h_{0a}^2\right)~. 
\end{align}
Let us recall that the constraint imposed by any
Lagrange multiplier term,
say $q^L C(q') $ has a double effect: (i) it eliminates some variable (possibly to be chosen
among several) entering the corresponding constraint equation, $C(q')=0$, and (ii) solving the latter constraint furthermore eliminates $q^L$ from the action. 

At the linear level, the term ${Q}^L_a C^{LO}_a$ imposes that $q_a=0$ and also eliminates ${Q}^L_a$. Then the term ${Q}^L_{[ab]} C^{LO}_{[ab]}$ imposes that $\check{Q}_{[ab]}=0$,
and also eliminates ${Q}^L_{[ab]}$. Finally, the
scalar term $ h_{00} C^{LO}$ sets the spatial trace $\bar h$ of $h_{ab}$ to zero, and also eliminates $h_{00}$. Those eliminations imply
also a simplification of the kinetic terms of the action. 
In view of the constraint on
$\bar h= \delta^{ab} h_{ab}$, it will be henceforth
convenient to generally decompose $h_{ab}$
in trace-free part ($h_{\langle ab\rangle} \equiv h_{ab} - \frac13 h_c^c \delta_{ab}$),
and trace part, namely
\begin{equation}
 h_{ab} = h_{\la ab\ra } + \frac13   \bar h \delta_{ab}\,,
\end{equation}
and similarly, with $\bar q \equiv \delta^{ab} q_{ab}$,
\begin{equation}
 q_{ab} = q_{\la ab\ra } + \frac13   \bar q \delta_{ab}\,.
\end{equation}
With this notation the kinetic terms of the action become
\begin{equation}
h_{\la ab\ra }\dot{q}^{\la ab\ra } + \frac13 \bar{h} \dot{\bar{q}}
    +\frac{1}{2} h_{0a}  \dot{q_a}   \,. 
\end{equation}
The linear constraints $\bar h=0$ and $q_a=0$
then eliminate two of the kinetic terms in the action and lead to a
reduced (but not yet fully reduced) action of the form
\begin{align}
    \frac{\mathcal{L}^{\rm red}_1}{c_F} = h_{\la ab\ra }\dot{q}^{\la ab\ra }-\mathcal{H}^{\rm red}[h_{\la ab\ra }, q_{\la ab\ra }, \bar{q}, h_{0a}, \check{Q}_{abc} ]~,
\end{align}
where 
\begin{align}
-\mathcal{H}^{\rm red}[h_{\la ab\ra }, q_{\la ab\ra }, \bar{q}, h_{0a}, \check{Q}_{abc} ] &=\frac{1}{6}\bar{q}^2- q_{\la ab\ra }^2+\check{Q}_{abc} \check{Q}_{cba}\nonumber\\
&-\frac{\kappa^2}{4}\left(-2h_{0a}^2  + h_{\la ab\ra }^2\right)~.
\end{align}
At this stage, the variables $\bar{q}$, $h_{0a}$ and $\check{Q}_{abc}$ enter only algebraically in the action, so that their respective equations of motion,
\begin{align}
&\frac{\partial \mathcal{H}^{\rm red}}{\partial \bar{q}  }=0, \\ \nonumber
&\frac{\partial \mathcal{H}^{\rm red}}{\partial h_{0a}  }=0, \\ \nonumber
&\frac{\partial \mathcal{H}^{\rm red}}{\partial \check{Q}^*_{ab}   }=0,
\end{align}
yield the additional constraints
\begin{equation}
\bar{q}=0,\quad h_{0a}=0,  \quad {\rm and} \quad \check{Q}^*_{ab} =0~.  
\end{equation}

Finally, the fully reduced linear action involves
only $h_{\la ab\ra }$, and $q_{\la ab\ra }$ and reads
\begin{align}
    \frac{\mathcal{L}^{\rm full \; red}_1}{c_F} = h_{\la ab\ra }\dot{q}^{\la ab\ra }-\mathcal{H}^{\rm full \; red}[h_{\la ab\ra }, q_{\la ab\ra }]~,
\end{align}
with 
\begin{equation}
-\mathcal{H}^{\rm full\; red}[h_{\la ab\ra }, q_{\la ab\ra }] = -\frac{\kappa^2}{4} h_{\la ab\ra }^2 - q_{\la ab\ra }^2\,.
\end{equation}
The above quadratic-in-fields Hamiltonian action
explicitly displays the presence of the five dofs
(i.e. five $q$'s and five $p$'s) of a linear
Fierz-Pauli (spatially homogeneous) massive spin-2 field, namely $h_{\la ab\ra }$ and its
canonical conjugate $q_{\la ab\ra }$ .

In the next section, we consider the nonlinear dynamics of torsion bigravity.

\section{Torsion bigravity at the nonlinear level}
\label{Sec:TGnon-linear}
At the nonlinear level, we still have seven first-class constraints associated with the seven Lagrange-multiplier terms  $h_{00} C +  {Q}^L_a C_a +\frac12 {Q}^L_{[ab]} C_{[ab]}$, but the corresponding constraints are less potent in eliminating dynamical variables than in the linear case, notably because they do not
eliminate kinetic terms. One way to see what are the remaining
number of dofs is to proceed as follows.

Having decomposed as above $h_{ab}$ and $q_{ab}$ in trace-free and trace parts, and using for later convenience
an integration by parts, we can rewrite the action so
that $q_a$ appears undifferentiated,
\begin{align}
    \frac{\mathcal{L}_1}{c_F} = h_{\la ab\ra }\dot{q}^{\la ab\ra } + \frac13 \bar{h} \dot{\bar{q}}
    -\frac{1}{2}  q_a \dot{h}_{0a} -\mathcal{H}~.
\end{align}
Introducing the following new list of variables
\begin{equation} \label{vars'}
    [{\rm vars'}]= [\bar{h}, q_a,\check{Q}_{[ab]}\;; \; \bar{q},h_{0a}, h_{\la ab\ra },q_{\la ab\ra }, 
    \check{Q}_{abc}]~,
\end{equation}
the Hamiltonian part of the action has the structure
\begin{align} \label{Hexact1}
-\mathcal{H} &= h_{00} C[{\rm vars'}] 
+  {Q}^L_a C_a[{\rm vars'}] \\ \nonumber
&+\frac12 {Q}^L_{[ab]} C_{[ab]}[{\rm vars'}] -\mathcal{H}^{\rm remain}[{\rm vars'}]~.
\end{align}

We separated the list of variables $[{\rm vars'}]$, Eq.\eqref{vars'}, in two blocks, the first block consists of the seven
variables
\begin{equation*}
   [\bar{h}, q_a,\check{Q}_{[ab]}]~.
\end{equation*}
This is because  we have seen above that, in the linear approximation, the seven Lagrange constraints 
are respectively proportional to:
$C^{LO} \propto \bar{h}$, 
$C^{LO}_a \propto q_a$,
$C^{LO}_{[ab]} \propto \check{Q}_{[ab]}$,
and thereby respectively eliminate these seven variables
$[\bar{h}, q_a,\check{Q}_{[ab]}] $.
At the nonlinear level, taking into account the $(1+p) q q + p^2 \sim (1+h) A A +h^2 $ structure of the Hamiltonian, the constraints
$C_a $, and $C_{[ab]}$ are deformed but
remain linear in $q_a$ and $\check{Q}_{[ab]}$
(with coefficients depending on $h_{\la ab\ra }$
and $\bar{h}$). One can then solve the linear system $C_a =0$, and $C_{[ab]}=0$ to get $q_a$ and $\check{Q}_{[ab]}$ as functions 
of $\bar{h}, h_{\la ab\ra }, q_{\la ab\ra }, h_{0a},  \bar{q},  \check{Q}_{abc}$. Inserting the latter
expressions in the scalar constraint
\begin{equation}
 0= C = - \kappa^2 \bar{h} + {\rm nonlinear \; terms}  ~,
\end{equation}
then yields a nonlinear equation for $\bar{h}$
which can, however, be perturbatively solved to express $\bar{h}$  as a function of the rest
of the list of variables of Eq.~\eqref{vars'}, namely
\begin{equation}
 \bar{q},h_{0a}, h_{\la ab\ra },q_{\la ab\ra }, 
    \check{Q}_{abc}   ~.
\end{equation}
The function $\bar{h}[{\rm vars'}]$ can be obtained
at any desired order as a power series in the remaining variables.

At this stage one has eliminated
$q_a$, $\check{Q}_{[ab]}$, and $\bar{h}$,
as well as their corresponding Lagrange multipliers,
${Q}^L_a$, ${Q}^L_{[ab]}$ and $h_{00}$.
This leads to a reduced action having the structure
\begin{align} \label{Lred1}
   & \frac{\mathcal{L}_1^{\rm red}}{c_F} = h_{\la ab\ra }\dot{q}^{\la ab\ra } +\frac13  \dot{ \bar{q}} \, \bar{h}^{\rm sol}[{\rm vars''};\check{Q}_{abc}] \\ \nonumber
  &  -\frac{1}{2}   \dot{h}_{0a} q_a^{\rm sol}[{\rm vars''};\check{Q}_{abc}] -\mathcal{H}^{\rm red}[{\rm vars''};\check{Q}_{abc}]~.
\end{align}
Here, the list of variables, $[{\rm vars''};\check{Q}_{abc}]$, entering
both the solutions obtained by solving the 
Lagrange constraints, $\bar{h}^{\rm sol}[{\rm vars''};\check{Q}_{abc}]$
and $q_a^{\rm sol}[{\rm vars''};\check{Q}_{abc}]$, as well as the nonkinetic
part of the action, $-\mathcal{H}^{\rm red}[{\rm vars''};\check{Q}_{abc}]$ comprises, as
indicated, two different sets of variables.
First, the reduced list
\begin{equation} \label{vars}
    [{\rm vars''}]= [\bar{q},h_{0a}, h_{\la ab\ra },q_{\la ab\ra }]\,,
\end{equation}
and second, the six variables $\check{Q}_{abc}$.
The reason why we separated $\check{Q}_{abc}$
from the list $[{\rm vars''}]$, is that the latter variables enter as dynamical variables in the kinetic terms of the action, while $\check{Q}_{abc}$,
 only appears algebraically in the action.

At this last stage we can  further
eliminate the six variables $\check{Q}_{abc}$
by  solving their (reduced) equation
of motion, namely (using the restricted components of $\check{Q}_{abc}$ to compute the derivative)
\begin{equation} \label{Q3constraint}
 0= c_F^{-1} \frac{\partial \mathcal{L}_1^{\rm red}}{\partial \check{Q}_{abc}} \,.
 \end{equation}
 This last constraint can be perturbatively solved
 for $\check{Q}_{abc}$ as a power series in the
 remaining dynamical variables because we have seen
 above that, at the linear level, $\check{Q}_{abc}$ appears quadratically in the
 Hamiltonian as a simple term $\propto \check{Q}_{abc} \check{Q}_{bca}$. Therefore, at the
 nonlinear level, the constraint \eqref{Q3constraint} will be of the form
 \begin{equation}
  0=\check{Q}_{bca} + {\rm nonlinear \; terms} \,. 
 \end{equation}
 Note, however, the following crucial new feature of
 the latter constraint at the nonlinear level:
 In the action \eqref{Lred1}, the six variables
 $\check{Q}_{abc}$ appeared not only in the reduced Hamiltonian $-\mathcal{H}^{\rm red}[{\rm vars};\check{Q}_{abc}]$
 but also in the kinetic terms
 through the couplings
 \begin{equation}
  +\frac13  \dot{ \bar{q}} \, \bar{h}^{\rm sol}[{\rm vars''};\check{Q}_{abc}]
    -\frac{1}{2}   \dot{h}_{0a} q_a^{\rm sol}[{\rm vars''};\check{Q}_{abc}]~.    
 \end{equation}
 As a consequence, the solution $\check{Q}^{\rm sol}_{abc}$ of the constraint Eq.~\eqref{Q3constraint} will depend not only on
 the list of variables $[{\rm vars''}]$, Eq.~\eqref{vars}, but also on $\dot{ \bar{q}}$
 and $\dot{h}_{0a}$. The elimination
 of  $\check{Q}_{abc}$ then leads to a
 final reduced action having the structure
 \begin{align} \label{Lred2}
   & \frac{\mathcal{L}_1^{\rm red'}}{c_F} = h_{\la ab\ra }\dot{q}^{\la ab\ra } +\frac13  \dot{ \bar{q}} \, \bar{h}^{\rm sol}[{\rm vars''};\dot{ \bar{q}}, \dot{h}_{0a}] \\ \nonumber
   & -\frac{1}{2}   \dot{h}_{0a} q_a^{\rm sol}[{\rm vars''};\dot{ \bar{q}}, \dot{h}_{0a}] -\mathcal{H}^{\rm red}[{\rm vars''};\dot{ \bar{q}}, \dot{h}_{0a}].
\end{align}
 This structure makes it clear that, at the nonlinear level, the reduced torsion bigravity action involves the time derivatives $\dot{ \bar{q}}$ and $\dot{h}_{0a}$ no longer in
 a linear way (\`a la $p \dot q$) but in a
 nonlinear way. 
 
 By explicitly performing the
 reduction process indicated above, we have
 perturbatively computed the final reduced action, Eq.~\eqref{Lred2}, as a power series
 in all the dynamical variables, starting
 at the quadratic level (corresponding to the
 linear dynamics), say
 \begin{equation}
   \frac{\mathcal{L}_1^{\rm red'}}{c_F} = \epsilon^2 + \epsilon^3 + \epsilon^4 + \cdots~,
 \end{equation}
 where $\epsilon$ is a counting parameter
  keeping track of the order in all the variables entering the final reduced action
 \begin{equation}
   \epsilon \sim  h_{\la ab\ra }, q_{\la ab\ra },\dot{q}_{\la ab\ra }; h_{0a},
      \dot h_{0a};\bar{q},\dot{\bar{q}}\,.
 \end{equation}
  In particular, we found that at order $\mathcal{O}(\epsilon^4)$, i.e. when considering terms
  quartic in the fields, there arise contributions {\it quadratic} in $\dot h_{0a}$
  with coefficients involving $ h_{\la ab\ra } $,
  e.g. terms of the form
  \begin{equation} \label{doth0asquare}
   \sim  h_{\la ac\ra }  h_{\la cb\ra }   \dot h_{0a} \dot h_{0b} \,.
  \end{equation}
  The presence of these terms means that the three variables $\dot h_{0a}$ which could
  be eliminated at the leading-order, $\mathcal{O}(\epsilon^2)$, introduce 
  three new dofs (beyond the normal five dofs
  described by the canonical variables $h_{\la ab\ra }, q_{\la ab\ra }$) at the quartic-in-field
  level. 
  
  At the quartic level, the variable $\bar{q}$
  appears only algebraically (but nonlinearly) in the reduced action. This implies that,
  at this level, the
  field equation of $\bar{q}$ would algebraically
  determine it in terms of the other variables,
  so that it would not represent a new dof.

  However, starting at the quintic level, 
  $\mathcal{O}(\epsilon^5)$, there are terms in the reduced
  action that couple $\dot{\bar{q}}$ to the
  time derivatives of $h_{0a}$, say terms
  of the structure
  \begin{equation}
  \sim   h_{\la ac\ra }  h_{\la cb\ra } \dot{\bar{q}} \, \dot h_{0a} \dot h_{0b} \,.
  \end{equation}
  When varying these terms with respect to $h_{0a}$ we see that the second time derivative of $\bar{q}$ will appear in the
  equations of motion. This shows that one
  now needs to give both $\bar{q}$ and $\dot{\bar{q}}$ as initial data. In other words, $\bar{q}$ enters the dynamics as a full additional dof at the quintic level. 
  In addition, at the sextic level, 
  $\mathcal{O}(\epsilon^6)$, there are terms in the reduced action that involve the square of 
  $\dot{\bar{q}}$, say terms of the form
  \begin{equation} \label{dotqsquare}
  \sim  h_{\la ac\ra }  h_{\la cb\ra } (\dot{\bar{q}})^2  \dot h_{0a} \dot h_{0b}\,. 
  \end{equation}
  The presence of such terms in the nonlinear action confirms that $\bar{q}$ provides a
  full additional dof in the nonlinear dynamics of torsion bigravity.

  \section{Conclusions} \label{Conclusions}

We studied the nonlinear regime of torsion gravity, a three-parameter class of modified gravity theories
involving a propagating (con)torsion $ K^i_{\; j\mu}=A^i_{\; j\mu}-\omega^i_{\; j\mu}$ in addition to the metric $g_{\mu \nu}$ (with associated Levi-Civita connection $\omega^i_{\; j\mu}$). We focussed
on the formal weak-gravity limit $16 \pi G \equiv \frac{1}{c_R} \to 0$ where $g_{\mu \nu}= \eta_{\mu\nu} + \frac{1}{\sqrt{c_R}} \hat h_{\mu \nu}$ decouples both from matter and from the dynamics of $A^i_{\; j\mu}$, and defines a Minkowski background  $\eta_{\mu\nu}=\eta_{ij}$. [Here, both Greek indices $\mu, \nu ,...$ and latin indices from the second part of the alphabet, $i,j,k,...$ take four values: $0,1,2,3$. ]

We studied the Poincar\'e-invariant nonlinear dynamics of the contorsion field $K_{[ij]k}=A_{[ij]k}$ in the decoupling limit. We showed that it could be described by several different actions, notably: (i)
a second-order action ${\mathcal L}_{TBG}[A_3]$, Eq. \eqref{Lagnew1}, involving only $A_{ijk}$;
(ii) a first-order action ${\mathcal L}_1[b_2,A_3]$,
Eq.~\eqref{eqn:SLag2xier}, involving both $A_{ijk}$ and a (Fierz-Pauli-like) symmetric 2-tensor $b_{ij}$; and (iii) a second-order action 
$\mathcal{L}_2 [b_2]$, Eq.~\eqref{eq:L2b}, involving only the Fierz-Pauli-like tensor $b_{ij}$. The latter field is related to the Ricci tensor $ {\mathcal F}_{ij}= F_{ij}[A_3]$ of the torsionfull connection $A_{ijk}$ via
\begin{equation} \label{bvsF2}
b_{ij}= \eta_{ij} + \frac{2}{\kappa^2} ({\mathcal F}_{ij} -\frac13 {\mathcal F} \eta_{ij}) \,.
\end{equation}
We also parametrize $b_{ij}$ by a metric-perturbation-like tensor $h_{ij}$ defined as
\begin{align}
    b_{ij}
    &\equiv \eta_{ij}+h_{ij}-\frac{1}{2}\eta_{ij}h\indices{_k^k}~.
\end{align}
The second-order action $\mathcal{L}_2 [b_2]$ defines a massive-gravity theory which differs from the usually considered ones in having a quadratic Fierz-Pauli mass term, but a non-Einsteinian kinetic term. We found convenient to study the nonlinear dynamics of the massive gravity theory $\mathcal{L}_2 [b_2]$ by means of the first-order action ${\mathcal L}_1[b_2,A_3]$,
Eq. \eqref{eqn:SLag2xier}.

We focussed on {\it spatially homogeneous} solutions of the latter torsion-inspired  massive gravity theory. Our final result is that  the dynamics defined by $\mathcal{L}_2 [b_2]$ involves
(in the spatially homogeneous case) {\it nine} dofs (instead of the five dofs of a
  normal, Fierz-Pauli massive spin-2 field).
Our analysis used the first-order action ${\mathcal L}_1[b_2,A_3]$ to display and solve the constraints implied by this action;
see Sec. \ref{Sec:TGnon-linear}. 

We define (with spatial indices $a,b,c,...=1,2,3$)
the torsion-related variable
\begin{align}
    q_{(ab)}=-A_{0(ab)}+\delta_{ab}A\indices{_{0c}^c}\,,   
  \end{align}
 and its spatial trace $\bar{q}= \delta_{ab} q_{(ab)} $, together with its trace-free part
 $q_{\la ab\ra } \equiv  q_{(ab)} -\frac13 \bar{q} \delta_{ab}$. Similarly,  $h_{\la ab\ra } \equiv  h_{ab} -\frac13 \bar{h} \delta_{ab}$, with 
 $\bar{h}= \delta_{ab} h_{ab} $.

The final reduced form of the action, after Legendre
  transforming the kinetic term 
  $h_{\la ab\ra } \dot q_{\la ab\ra } $   (which is the only occurrence of $\dot q_{\la ab\ra } $ 
  in the reduced action) 
 can be described by a {\it second-order Lagrangian}
  with the structure
\begin{align}
  \mathcal{L}\left(  h_{\la ab\ra },\dot h_{\la ab\ra }; h_{0a},
      \dot h_{0a};\bar{q},\dot{\bar{q}}\right) ~  ,
\end{align}
  where all the time-derivative terms appear nonlinearly, though at different orders in nonlinearity: $\dot h_{\la ab\ra }$ enters at
  the leading (Fierz-Pauli), quadratic order
  \begin{equation}
  \frac{\mathcal{L}^{LO}}{c_F}= \frac14 \left( (\dot h_{\la ab\ra })^2 - \kappa^2 (h_{\la ab\ra })^2 \right),
\end{equation}
while $\dot h_{0a}$ enters bilinearly at the quartic order, Eq. \eqref{doth0asquare}, and $\dot{\bar{q}}$ enters bilinearly at the sextic order, Eq. \eqref{dotqsquare}.

  The general solution of the
 corresponding second-order equations of motion then requires that one
  gives initial data for the eighteen variables
  \begin{equation}
  h_{\la ab\ra },\dot h_{\la ab\ra }; h_{0a},
      \dot h_{0a};\bar{q},\dot{\bar{q}}  \,,
\end{equation}
corresponding to the nine dofs $h_{\la ab\ra }, h_{0a}, \bar{q}$.

  It is possible that the appearance of the four new dofs (here parametrized by $h_{0a}$ and $\bar{q}$) would arise in a different way,
  and at lower perturbative orders, when considering generic,
  inhomogeneous solutions of torsion bigravity
  (notably via the presence of spatial gradients in the constraints). However, on the one hand,
  our simplifying  homogeneity ans{\"a}tz is consistent\footnote{As a side check, we applied the approach of homogeneous solutions to the usual massive-gravity theories (containing either a ``bad" nonlinear Fierz-Pauli mass term, or a ``good"  the deRham-Gabadze-Tolley one) and we
  confirmed that this approach predicts in
  all cases the correct counting of dof.}, and, on the other hand, it suffices to conclude that
  torsion gravity theories do not define a healthy theory containing, in all nonlinear situations, only the hoped-for $2 + 5$ dofs of   massless spin-2 and massive spin-2 excitations. 
  
  Our negative conclusion concerning the healthiness of torsion bigravity (and thereby of the more general 6-parameter torsion gravity theories) is not in conflict with the  positive results obtained 
  in previous works Refs.~\cite{Nair:2008yh,Nikiforova:2009qr,Deffayet:2011uk,Nikiforova:2018pdk,Damour:2019oru,Nikiforova:2020fbz,Nikiforova:2021xcj,Nikiforova:2022pqy} because the latter results had studied less general dynamical configurations. Our proof above has shown that the four additional dof were rather hidden,
  because they only start to arise 
  at the quartic order in torsion variables and
  in time-dependent situations.

  \section*{Acknowledgements}
  T.~J. thanks the hospitality and the stimulating environment of the Institut des Hautes Etudes Scientifiques. T. J. is supported by a LabEx Junior Research Chair Fellowship. The present research was also partly supported by the ``\textit{2021 Balzan Prize for 
Gravitation: Physical and Astrophysical Aspects}'', awarded to Thibault Damour. 
\bibliography{bibliography}
\end{document}